\documentclass[aps,prl,amsmath,amssymb,preprintnumbers,%
twocolumn,showpacs,floatfix]{revtex4}
\usepackage{epsfig,graphics}
\usepackage{graphicx}% Include figure files
\usepackage{dcolumn}% Align table columns on decimal point
\usepackage{bm}% bold math
\usepackage{amsmath}
\usepackage[usenames]{color}

\newcommand{\ave}[1]{\left\langle #1 \right\rangle} %regular average
\newcommand{\third}{\mbox{${1\over3}$}}         %       1/3

\begin{document}

\title{Baryon-strangeness correlations:
a diagnostic of strongly interacting matter}

\author{V.\ Koch, A.\ Majumder, and J.\ Randrup}

\affiliation{Nuclear Science Division,
Lawrence Berkeley National Laboratory,
1 Cyclotron road, Berkeley, CA 94720}

\date{\today}

\begin{abstract}
The correlation between baryon number and strangeness 
%is proposed as a diagnostic for 
elucidates the nature of strongly interacting matter,
such as that formed transiently in high-energy nuclear collisions.
This diagnostic %correlation 
can be extracted theoretically from lattice QCD calculations
and experimentally from event-by-event fluctuations.
The analysis of present lattice results above the critical temperature
severely limits the presence of $q\bar q$ bound states,
thus supporting a picture of independent (quasi)quarks.
\end{abstract}

	%	Particle correlations

\preprint{LBNL-57613R}

\maketitle

%=======================================================================
The principal goal of high-energy heavy-ion collisions 
is the creation and exploration of a novel state of matter
in which the quarks and gluons are deconfined over distances 
considerably larger than that of a hadron \cite{Collins:1974ky}. 
It had originally been assumed that asymptotic freedom would cause
such matter to behave as a plasma of massless quarks and gluons
interacting with a relatively weak screened chromodynamic Coulomb force.
While this picture is supported qualitatively
by the rapid rise in the entropy density 
at a temperature of about $T_c \simeq 170~{\rm MeV}$,
as obtained by lattice QCD calculations,
the fact that the high-$T$ behavior falls somewhat below 
that of an ideal gas of massless quarks and gluons \cite{Karsch:2003jg},
indicates that the chromodynamic plasma has a more complex structure.

Indeed, recent results from lattice QCD calculation on spectral functions
\cite{Hatsuda_1,Hatsuda_2,Hatsuda_3,Karsch_spectral} 
suggest the presence of bound, 
color-neutral states well above $T_c$. 
This has led to the suggestion that at moderate temperatures, 
$T \simeq 1-2 \,T_c$,
the system is composed of medium-modified (massive) quarks and gluons
together with their (many and possibly colored)
 bound states \cite{Shuryak:2003ty,Shuryak:2004tx,Shuryak:2004cy,GEB}. 

Furthermore,
jet quenching observed at the Relativistic Heavy-Ion Collider (RHIC) 
suggests that the matter formed is not hadronic \cite{quench}.
Moreover, the RHIC collisions exhibit strong elliptic and radial 
collective flows that are consistent with predictions by ideal fluid dynamics
\cite{Ackermann:2000tr,Adler:2003kt}.
If ideal hydrodynamics is indeed the correct framework, 
rapid thermalization must occur
and this in turn would seem to rule out a weakly interacting plasma.
Thus, the nature of the matter being created at RHIC is still not clarified
and the purpose of the present Letter is to offer a novel diagnostic tool
for eluciding this issue by
probing the relevant degrees of freedom and their correlations.

In particular we will argue that 
the correlation between the strangeness $S$ and the baryon
number $B$ provides a useful diagnostic for the presence of 
strong correlations between quarks and anti-quarks.
In order to understand this,
consider first a situation in which the basic degrees of freedom
are weakly interacting quarks and gluons. 
Then strangeness is carried exclusively by the $s$ and $\bar s$ quarks
which in turn carry baryon number in strict proportion to their strangeness,
$B_s=-\frac{1}{3}S_s$,
thus rendering strangeness and baryon number strongly correlated. 
This feature is in stark contrast to a hadron gas 
in which the relation between strangeness and baryon number is less intimate.
For example, at small baryon chemical potential
the strangeness is carried primarily by kaons, which have no baryon number.

These elementary considerations lead us to introduce the
following correlation coefficient,
\begin{equation}
C_{BS} \equiv -3{\sigma_{BS}\over\sigma_S^2} =
-3 \frac{\ave{BS}-\ave{B}\ave{S}}{\ave{S^2} - \ave{S}^2}
= -3 \frac{\ave{BS}}{\ave{S^2}} .
\end{equation}
The average $\ave{\cdot}$ is taken over a suitable ensemble of events
and the last expression uses the fact that $\ave{S}$ vanishes.
When the active degrees of freedom are individual quarks,
the total strangeness is $S=\nu_{\bar s}-\nu_s$,
while the baryon number can be expressed as $B=\third(U+D-S)$,
where $U=\nu_u-\nu_{\bar u}$ is the upness
and   $D=\nu_d-\nu_{\bar d}$ is the downness.
Thus, if the flavors are uncorrelated, we have
$\sigma_{BS}=-\third\sigma_S^2$ and $C_{BS}$ is unity.

\begin{figure}[htbp]
\includegraphics[width=3.2in]{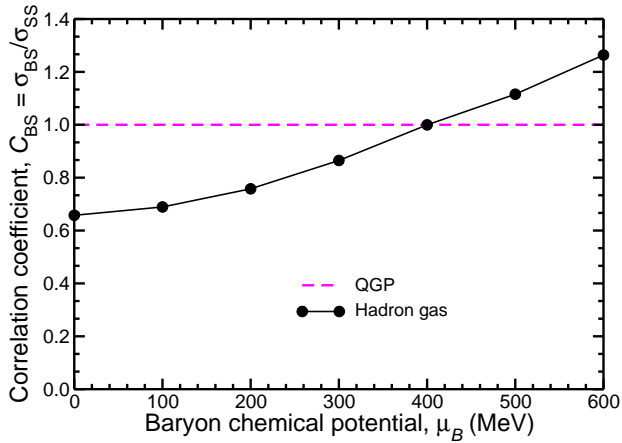}
\caption{The correlation coefficient $C_{BS}=\sigma_{BS}/\sigma_S^2$
for a hadron gas (including all species up to the $\Omega^-$)
at freeze-out, shown as a function of the baryon chemical potential $\mu_B$
(with the temperature $T$ decreasing from 170~MeV as $\mu_B$ is increased,
as found in Ref.\ \cite{Braun-Munzinger:2003zd}).
Also shown is the corresponding result 
for an ideal quark-gluon plasma (dashed line).
}\label{fig1}
\end{figure}

In a gas of hadron resonances,
the total baryon number is $B=\sum_k n_k B_k$
and its total strangeness is $S=\sum_k n_k S_k$,
where the specie $k$ has baryon number $B_k$ and strangeness $S_k$.
If no prior correlations are present,
the correlation coefficient may then be expressed in terms of the multiplicity 
variances $\sigma_k^2\equiv\ave{n_k^2}-\ave{n_k}^2\approx\ave{n_k}$,
\begin{equation}
C_{BS}\ =\ -3\,\frac{\sum_{k} \sigma_k^2 B_k S_k}{\sum_{k} \sigma_k^2 S_k^2}\
\approx\ -3\,\frac{\sum_{k} \ave{n_k} B_k S_k}{\sum_{k} \ave{n_k} S_k^2}\ ,
\end{equation}  
where the approximate expression holds for Poisson statistics.
Thus, in the hadronic gas,
the numerator receives contributions
from only (strange) baryons (and anti-baryons),
while the denominator receives contributions also from (strange) mesons,
\begin{equation}
C_{BS}\ \approx\ 3\,{\ave{\Lambda} + \ave{\bar{\Lambda}} + \dots\,
+3 \ave{\Omega^-} + 3\ave{\bar{\Omega}^+} \over
K^0 + \bar{K}^0 + \dots\,
+9 \ave{\Omega^-} + 9\ave{\bar{\Omega}^+}}\ .
\end{equation} 
The hadronic freeze-out value value of $C_{BS}$ is shown in Fig.\ \ref{fig1}. 
At the relatively high temperatures relevant at RHIC,
the strange mesons significantly outnumber the strange baryons,
so $C_{BS}$ is smaller than unity.
Indeed, including hadrons up to $\Omega^-$,
we find $C_{BS}=0.66$ for $T=170~{\rm MeV}$ 
and zero chemical potential, $\mu_B=0$.
As $\mu_B$ is increased, the freeze-out temperature decreases
\cite{Braun-Munzinger:2003zd} and, consequently, 
strange baryons steadily gain favor relative to antikaons, 
so $C_{BS}$ increases.
As this trend continues, only very few antibaryons are present
so the (positive) strangeness is carried primarily by kaons 
and compensated by $\Lambda$ and $\Sigma$ hyperons.
We then have $\ave{\Lambda+\Sigma}\approx\ave{K}$
and so $C_{BS}\approx\frac{3}{2}$.
This significant dependence of $C_{BS}$ on the hadronic environment
is in sharp contrast to the simple quark-gluon plasma
where the correlation coefficient remains strictly unity
at all temperatures and chemical potentials.

%.......................................................................
%.......................................................................

Whether the chromodynamic system,
with respect to its baryon-strangeness correlations,
does in fact appear as an assembly of elementary quasiparticles 
can be checked by lattice calculations.
We thus note that in statistical equilibrium
the correlation coefficient can be expressed as a ratio of susceptibilities,
$C_{BS}=-3\chi_{BS}/\chi_{SS}$,
which in turn are second derivatives of the free energy
with respect to the chemical potential(s),
\begin{equation}
\chi_{BS} = -\frac{1}{V} \frac{\partial^2 F}{\partial\mu_B\partial\mu_S}\ ,\
\chi_{SS} = -\frac{1}{V} \frac{\partial^2 F}{\partial\mu_S^2 }\ .
\end{equation}
In terms of the basic flavor densities $u, d, s$ we thus have
\begin{equation}
C_{BS}\ =\ 1 + \frac{\chi_{ds} + \chi_{us} }{ \chi_{ss} }\ .  
\end{equation}
At $\mu_B=0$,   %zero baryon density, 
the mixed flavor susceptibilities generally tend to be relatively small
above $T_c$ \cite{AlltonPRD71}.
Indeed, using the values $\chi_{ff'}$ extracted at $T=1.5\,T_c$
by Gavai {\em et al.} \cite{Gavai:2002jt}, %OR: ,Gavai:2002kq}, 
we obtain $(\chi_{us}+\chi_{ds})/\chi_{ss} = 0.00 (3)/0.53 (1) \ll1$.
These results were obtained in a quenched approximation, 
but the effect of sea quarks is expected to be marginal above $T_c$
\cite{Gavai:2002jt}.
We may thus surmise that the lattice system has $C_{BS}\approx1$,
suggesting that the quark flavors are uncorrelated,
as in the ideal quark-gluon plasma.
(However, the presence of pure gluon clusters cannot be ruled out
by this diagnostic.)
It would obviously be of interest to 
move the lattice calculations beyond the quenched approximation.

Since the ratio of strange to non-strange susceptibilities,
$\chi_{ss}/(\chi_{uu} + \chi_{dd})$, obtained from lattice QCD 
agrees with that of a hadron gas
right at the critical temperature $T_c$ \cite{Gavai:2002kq},
it would be interesting to see what lattice QCD
would yield for $C_{BS}$ at this temperature.
Unfortunately, to our knowledge,
no off-diagonal susceptibilities involving strange quarks
have yet been calculated.
However, the available results for light quarks \cite{AlltonPRD71}
indicate that the off-diagonal susceptibilities 
are smaller than the diagonal ones by a factor of twenty at $T_c$. 
Thus, if flavor symmetry holds, 
the lattice value of $C_{BS}$ would differ from
that of the hadron gas already at $T_c$.
We also note that the vanishing of the off-diagonal susceptibilities,
and hence the unit value of $C_{BS}$,
does not conflict with the existence of 
hadron-like resonances that have been identified well above $T_c$
\cite{Hatsuda_2,Hatsuda_3,Karsch_spectral},
since their large masses (of more than 2~GeV)
make them insignificant near $T_c$.

It is important to recognize that 
both the calculated equation of state \cite{Karsch:2003jg}
and the observed collective flow \cite{Ackermann:2000tr,Adler:2003kt}
indicate that the system cannot be merely an assembly of 
weakly interacting elementary quarks and gluons.
However, the various apparently conflicting features might be reconciled
if the system were to organize itself into an assembly of weakly interacting
quasiparticles, such as the picture emerging from QCD 
by the application of resummation techniques \cite{Blaizot}.

Recently there has appeared a model
that purports to explain both the equation of state 
as obtained on the lattice as well as the 
large flow observed in heavy-ion collisions 
\cite{Shuryak:2003ty,Shuryak:2004tx,Shuryak:2004cy}. 
The model describes the chromodynamic system
as a gas of massive quarks, antiquarks, and gluons together with
a myriad of their bound states generated by a screened Coulomb potential.
In order to assess the consistency of this model with present lattice results,
we estimate the ratio $C_{BS}$ in such a scenario.

We base our estimates on Ref.~\cite{Shuryak:2004tx} in which 
the temperature dependence of the screening length and the effective masses
were obtained by parametrizing lattice results.
The resulting attraction produces a total of 749 bound states,
of which only the color-triplet $sg$ and the color-singlet $q\bar s$ states
(and their conjugates) are of relevance here.
(The color-hexaplet $sg$ states as well as the diquark states
are very weakly bound and dissolve entirely at  the temperature $1.5\,T_c$
considered here).
There are 4 $\pi$-like (spin-singlet)
and 12 $\rho$-like (spin-triplet) $q\bar s$ states
as well as 18 $sg$ states (and their conjugates).
The abundancies of these states are estimated in a grand canonical ensemble
with vanishing chemical potentials.
The $q\bar s$ multiplets carry no baryon number 
and thus contribute only to $\sigma_S^2$,
hence drive $C_{BS}$ towards zero,
while the more degenerate $sg$ multiplet contributes also 
to $\sigma_{BS}$ and thus drives $C_{BS}$ towards unity.
The resulting value obtained at $T=1.5T_c$ is $C_{BS}=0.62$.
The fact that it differs significantly from the value extracted 
directly from lattice QCD (see above)
suggests that the bound-state model, in spite of its intuitive appeal, 
cannot account for the $BS$ correlations implied by the underlying QCD theory.
(No particular significance should be attached to the fact that
 the value of $C_{BS}$ obtained in the bound-state scenario
comes out as fairly close to the value of a standard hadron gas.)

\begin{figure}[htbp]
\includegraphics[width=3.2in]{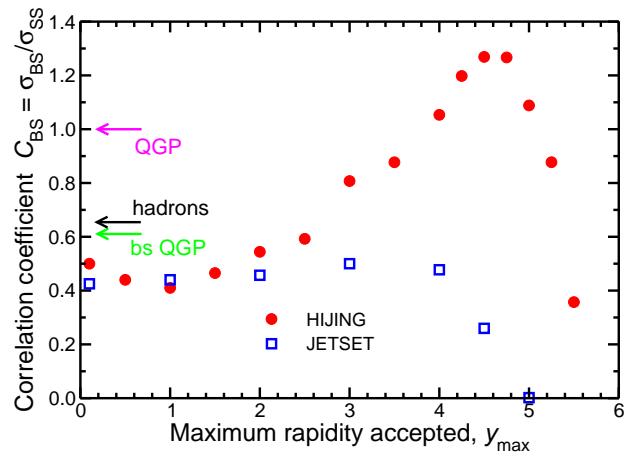}
\caption{The correlation coefficient $C_{BS}$ 
obtained with HIJING \cite{Wang:1991ht} 
from generated samples of 10,000 Au+AU events 
at the top RHIC energy of $\sqrt{s}=200\,A\,{\rm GeV}$,
shown as a function of the maximum rapidity accepted, $|y|\leq y_{\rm max}$.
(The points scatter due to the limited event statistics.)
Also shown is the corresponding results for $e^+e^-$ 
at $\sqrt{s}=200\,{\rm GeV}$ calculated with JETSET 
\cite{Andersson:1983ia}.
The arrows point to the values obtained for
the ideal quark-gluon plasma,
the hadronic gas at $T=170~{\rm MeV}$, and 
the bound-state QGP advocated in Ref.\ \cite{Shuryak:2004tx}
(at $1.5\,T_c$).
}\label{fig2}
\end{figure}

In order to estimate
 what to expect for a RHIC event if {\em no} QGP is generated,
we have extracted $C_{BS}$ from events generated by HIJING \cite{Wang:1991ht}
which presents an extension of the string fragmentation treatment
of hadron production in $pA$ and $AA$ collisions.
Thus we calculate
\begin{equation}
C_{BS}\ =\ -3\,{\sum_n B^{(n)} S^{(n)} - 
\frac{1}{N}\left(\sum_n B^{(n)}\right) \left(\sum_n S^{(n)}\right) 
\over \sum_n (S^{(n)})^2 
        - \frac{1}{N}\left(\sum_n S^{(n)}\right)^2}\ ,
\end{equation}
where $B^{(n)}$ and $S^{(n)}$ denote the baryon number and strangeness
observed for a given event $n$ within a central rapidity interval,
$|y|<y_{\rm max}$.
The result for Au+Au at the top RHIC energy of $100,A\,{\rm GeV}$
is shown in Fig.\ \ref{fig2}
as a function of the rapidity acceptance range $y_{\rm max}$.
At small values of $y_{\rm max}$, 
only hadrons emerging within a narrow slice at mid rapidity are included
and in the limit where no event contributes more than one strange hadron
the resulting value of $C_{BS}$ equals
the ratio of central rapidity density of baryonic strangeness 
to the central rapidity density of all strangeness
(The fact that the result, $C_{BS}\approx0.50$, lies somewhat below the value 
shown in Fig.\ \ref{fig1} for $\mu_B=0$ reflects the general underproduction
of strange baryons by the string mechanism.)
As the acceptance is increased, a broader range of chemical potentials
are sampled and, as a result, the value of $C_{BS}$ goes up,
following qualitatively the behavior with $\mu_B$ 
exhibited in Fig.\ \ref{fig1}. 
Finally, when $y_{\rm max}$ is increased towards the point 
where all baryons are accepted in each event,
the baryon fluctuation disappears and, consequently, $C_{BS}$ tends to zero
(while the strangeness still exhibits a small fluctuation due to weak decays).

To illustrate the effect of this canonical suppression effect,
we include results obtained for $e^+e^-$ collisions 
at $\sqrt{s}=200~{\rm GeV}$ with the underlying JETSET code 
\cite{Andersson:1983ia},
for which the net baryon density vanishes at all rapidities.

As a further reference, the correlation coefficient
has been exctracted from RHIC events
generated by RQMD \cite{NuXu}, yielding $C_{BS}\approx0.55$.
(This value is also lower than that associated with the hadronic freeze-out,
reflecting the underestimate of the baryonic strangeness by RQMD.)

%.......................................................................
%.......................................................................

While it is relatively straightforward to extract $C_{BS}$ 
from the various models, as brought out above,
it is more difficult to determine it experimentally.
In principle, $C_{BS}$ can be obtained from event-by-event fluctuation 
analysis designed to measure susceptibilities \cite{Jeon:2003gk}. 
Since both baryon number and strangeness are conserved quantities 
their correlations and fluctuations should be preserved during 
the expansion stage, 
provided the kinematic acceptance is large compared to 
the typical smearing associated with the hadronization. 
At the same time it should be sufficiently small
to render global conservation effects negligible.
This situation is analogous to the analysis of net charge fluctuations 
made in Refs.\ \cite{Jeon:2000wg,Asakawa:2000wh,Bleicher:2000ek}.

Relative to that work, the present situation is in fact 
somewhat more favorable,
since the principal carriers of baryon number and strangeness 
are heavier than those of charge (pions),
so one would expect smaller kinematic shifts during the hadronization. 
Furthermore, strong decays of strange baryons into kaons and nucleons, 
which in principle could wash out the correlation 
when small acceptances are used, are relatively rare \cite{gadziki}. 
On the other hand, an actual experimental investigation should
correct for the occurrence of weak decays, which remove strangeness,
and also address the problem of neutron detection. 
These issues remain to be studied in detail. 

In this Letter, we have proposed the correlation coefficient
$C_{BS} = -3\sigma_{BS}/\sigma_S^2$
as a useful diagnostic tool for elucidating the character
of chromodynamic matter.
If the baryon number and strangeness attributes are carried by
effectively independent quarks and antiquarks,
then this ratio is unity 
for any value of the energy and baryon-number density.
By contrast, it exhibits a significant dependence on the baryon density
in a hadron gas and is about $\frac{2}{3}$ at RHIC.
We have pointed out that the value for $C_{BS}$ obtained
from (quenched) lattice QCD calculations are close to unity,
while the recently proposed ``bound-state quark-gluon plasma'' 
yields significantly lower values.
We also noted that the commonly employed event generators
HIJING and RQMD yield $C_{BS}$ values at RHIC 
that are even lower than those associated with the hadronic freeze-out.
Thus $C_{BS}$ appears to provide a useful tool for distinguishing between
effective strangeness-carrying degrees of freedom in various models.

The current experimental activity is concentrated at RHIC
and we have therefore considered scenarios having vanishing chemical potential,
which are also easier to treat with lattice QCD.
However, since the hadronic $C_{BS}$ exhibits a marked dependence on $\mu_B$,
it would be interesting to explore this observable over a wide range
of net baryon densities and thus broaden the insight into the
character of the matter formed.

Finally, we recall that lattice QCD matter
has an equation of state that falls below 
that of a free gas of quarks and gluons at high temperature
while nevertheless being devoid of flavor correlations
(as reflected in the vanishing mixed flavor susceptibilities
and causing the strangeness correlation coefficient $C_{BS}$ to be unity).
We wish to suggest that these apparently inconsistent properties
might be reconciled if strongly interacting matter above $T_c$ would
organize itself into effectively independent but massive quasiparticles.
The fact that the lattice equation of state can be reproduced
in such a quasiparticle description was already demonstrated
in the $ud$ sector \cite{kaempfer}.
We emphasize again, however,
that the proposed diagnostic is insensitive to the possible existence
of strong correlations among the gluons
(whose presence could lead to a hydrodynamic behavior).

The reduction of the complicated many-body problem 
to the relative simplicity of the quasi-particle picture
is analogous to the highly successful 
independent-particle description of atomic nuclei
and could conceivably emerge from the dressing of 
the elementary quarks and gluons by their mutual interaction.
In such a picture the effective masses arise 
from a predominantly repulsive interaction 
which in turn would cause a collective expansion
and thus possibly help explain the observed flow in RHIC experiments.\\

The authors wish to thank S.\ Huang and N.\ Xu for providing us
with RQMD results for the correlation coefficient
and to also thank X.N.\ Wang for help with the HIJING code.
This work was supported by the Director, Office of Science, 
Office of High Energy and Nuclear Physics, 
Division of Nuclear Physics, and by the Office of Basic Energy
Sciences, Division of Nuclear Sciences, of the U.S. Department of Energy 
under Contract No. DE-AC03-76SF00098.

%=======================================================================

%=======================================================================

\end{document}